\documentclass{article}

\usepackage{arxiv}

\usepackage[utf8]{inputenc} % allow utf-8 input
\usepackage[T1]{fontenc}    % use 8-bit T1 fonts
\usepackage{hyperref}       % hyperlinks
\usepackage{url}            % simple URL typesetting
\usepackage{booktabs}       % professional-quality tables
\usepackage{amsfonts}       % blackboard math symbols
\usepackage{nicefrac}       % compact symbols for 1/2, etc.
\usepackage{microtype}      % microtypography
\usepackage{lipsum}		% Can be removed after putting your text content
\usepackage{graphicx}
\usepackage{natbib}
\usepackage{doi}
\usepackage{multirow}

\title{Development Of Automated Cardiac Arrhythmia Detection Methods Using Single Channel ECG Signal}

\author{ Arpita Paul\\
	Department of Electronics and Telecommunication \\Engineering,
	IIEST, Shibpur, West Bengal, India \\
	\texttt{paularpita1998@gmail.com} \\
	\And
	Avik Kumar Das \\
	Department of Electronics and Telecommunication \\Engineering, 
	IIEST, Shibpur, West Bengal, India \\	
	\texttt{avik.rs2018@telecom.iiests.ac.in } \\
 \And
	Manas Rakshit \\
	Department of Electronics and Communication Engineering\\
	NIT, Agartala, Tripura, India \\
		\texttt{manasece@nita.ac.in} \\
 \And
	Ankita Ray Chowdhury \\
	Department of Electronics and Telecommunication \\Engineering,
	IIEST, Shibpur, West Bengal, India\\
	\texttt{ankitarc.rs2020@telecom.iiests.ac.in} \\
 \And
	Susmita Saha \\
	Department of Electronics and Telecommunication \\Engineering,
	IIEST, Shibpur, West Bengal, India \\
	\texttt{2021etm001.susmita@students.iiests.ac.in} \\
  \And
	Hrishin Roy \\
	Department of Electronics and Telecommunication \\Engineering,
	IIEST, Shibpur, West Bengal, India \\
	\texttt{2022etm003.hrishin@students.iiests.ac.in} \\
 \And
	Sajal Sarkar \\
	Department of Electronics Science\\
	Seth Anandram Jaipuria College, Kolkata, India \\
	\texttt{sajalsarkar555@gmail.com} \\
 \And
	Dongiri Prasanth \\
	Department of Mining Engineering\\
	IIEST, Shibpur, West Bengal, India \\
	\texttt{prashanthdongiri95@gmail.com} \\
  \And
	Eravelli Saicharan \\
	Department of Mining Engineering\\
	IIEST, Shibpur , West Bengal, India\\
	\texttt{charanraoeravelli007@gmail.com} \\
}

\renewcommand{\shorttitle}{\textit{arXiv} Template}

\hypersetup{
pdftitle={A template for the arxiv style},
pdfsubject={q-bio.NC, q-bio.QM},
pdfauthor={David S.~Hippocampus, Elias D.~Striatum},
pdfkeywords={First keyword, Second keyword, More},
}

\begin{document}
\maketitle

\begin{abstract}
Arrhythmia, an abnormal cardiac rhythm, is one of the most common types of cardiac disease. Automatic detection and classification of arrhythmia can be significant in reducing deaths due to cardiac diseases. This work proposes a multi-class arrhythmia detection algorithm using single channel electrocardiogram (ECG) signal. In this work, heart rate variability (HRV) along with morphological features and wavelet coefficient features are utilized for detection of 9 classes of arrhythmia. Statistical, entropy and energy-based features are extracted and applied to machine learning based random forest classifiers. Data used in both works is taken from 4 broad databases (CPSC and CPSC extra, PTB-XL, G12EC and Chapman-Shaoxing and Ningbo Database) made available by Physionet. With HRV and time domain morphological features, an average accuracy of 85.11\%, sensitivity of 85.11\%, precision of 85.07\% and F1 score of 85.00\% is obtained whereas with HRV and wavelet coefficient features, the performance obtained is 90.91\% accuracy, 90.91\% sensitivity, 90.96\% precision and 90.87\% F1 score. The detailed analysis of simulation results affirms that the presented scheme effectively detects broad categories of arrhythmia from single-channel ECG records. In the last part of the work, the proposed classification schemes are implemented on hardware using Raspberry Pi for real time ECG signal classification.
\end{abstract}

% keywords can be removed

\section{Introduction}\label{sec:introduction}
The group of heart disorders, commonly called cardiovascular diseases (CVDs), have become the cause of an increasing number of premature deaths worldwide \cite{tsao2022heart}. Since 1990, prevalent cases of CVDs all over the world have doubled and the number of deaths due to the same has increased from 12.1 million in 1990 to 18.6 million in 2019 \cite{b1}. According to the world health organization (WHO), 32 \% of all global deaths in 2019 were from CVDs \cite{b2}. Over three-quarters of these deaths took place in poor countries with low doctor-to-patient ratios and inadequate medical infrastructure. Efficient and automated computer-aided diagnosis of CVDs can significantly reduce the burden on already strained healthcare systems, leading to timely detection and treatment of patients that results in reducing the number of deaths due to chronic heart diseases \cite{faust2016computer}.

ECG is the most extensively used non-invasive method for the clinical detection of CVDs \cite{hong2022practical}. ECG signal is difficult to analyze visually for long-term application due to its non-stationary nature. That may lead to missed or late diagnosis of life-threatening heart ailments. Automated signal processing with a machine learning algorithm can be developed to process ECG data in real-time for accurate and prompt detection of cardiac diseases with less human effort and error \cite{bertsimas2021machine}, \cite{b33}.

Arrhythmia is a heart condition characterized by an irregular heart rate where the heart beats either too slow or too fast. It occurs due to improper electrical impulses that coordinate the heartbeats. Rhythms are classified into different categories as per their origin like sinus rhythm, atrial rhythm, atrioventricular (AV) node rhythm, ventricle rhythms etc. Each category of rhythms is further sub-divided into several classes as per characteristics of rhythms \cite{b3}.

Methods for diagnosing arrhythmia using classification approaches based on a variety of ECG parameters have been proposed in a number of promising research \cite{b16,b17,b18,b19,b20,b21,b22}. Most of the works have classified individual beats into arrhythmia classes. Some works have focused on single arrhythmia type detection \cite{b16},\cite{b17} while other works have classified more than one class of arrhythmia  \cite{b18,b19,b20,b21,b22}. In \cite{b16}, relative wavelet energy from wavelet decomposition of T-Q segments in the ECG cycle is used for the detection of atrial fibrillation (AF). In \cite{b17}, an AF detection method is proposed using features from Stationary Wavelet Transform (SWT) decomposition coefficients and support vector machine as the classifier. The technique described in \cite{b18}, offers training several classifiers to distinguish between premature atrial contraction (PAC), premature ventricular contraction (PVC), and normal beats based on the R-R interval and other statistical features.  In \cite{b19}, distinguishing between ventricular fibrillation and ventricular tachycardia is done using time-frequency representations of ECG signal. The method in \cite{b20} recognizes and categorizes four groups of ECG beats using a set of nonlinear features, including Shannon entropy, fuzzy entropy, approximate entropy, permutation entropy, sample entropy, etc. When analyzing patient-specific based arrhythmia classification, the method described in \cite{b21} employs self-organized operational neural networks to detect and categorize five groups of arrhythmia beats. In \cite{b22}, five categories of heartbeat classification have been done using wavelet transform coefficients and independent component analysis (ICA).

Most of the previous works in the literature have used either only one or very limited databases and the performance of such methods are tested only in a small and homogeneous population. Moreover, the classification ability of the existing approaches is limited to only very few rhythm types. Broad-range classification of multiple classes has not been attempted much in the available literature. Majority of the works in arrhythmia classification have focused on classifying single heartbeat classification only, not arrhythmic episode detection. In doing so, a significant amount of inter-beat information is lost which restricts the classification to only a few classes of arrhythmia. Considering the above-mentioned aspect, the primary aim of this work is to propose a multi-class rhythm classification scheme to detect arrhythmia from ECG records in widely collected different unbalanced databases. The contributions of the paper are as follows:

\begin{itemize}
	
	\item  Development of robust multi-class ECG arrhythmia classification schemes which can effectively detect broad categories of arrhythmia like normal sinus rhythm (NSR), sinus arrhythmia (SA), sinus bradycardia (SB), sinus tachycardia (STACH), atrial fibrillation (AF), atrial flutter (AFL), premature atrial contraction (PAC), 1st Degree AV block (1AVB) and premature ventricular contraction (PVC).

     \item  Utilization of broadly acquired ECG records combined from four standard databases of Physionet challenge 2021 which makes the model more robust to different methods of data acquisition and population demography.
     
     \item Extraction of robust heart rate variability and time domain features based on scientific knowledge of characteristics of ECG signal variations caused by each of the arrhythmia classes. Also, deriving the feature set from established facts used for actual pathological diagnosis by physicians which provides the classifier model interpretability.
	
	\item Extraction of effective features using SWT-based sub-band signal decomposition. Deployment of heart rate variability features along with wavelet coefficient features for efficient classification of multi-class cardiac arrhythmia. 

    \item Hardware implementation of the ECG classification algorithms using Raspberry Pi for real time ECG signal analysis.
	
\end{itemize}

The rest of the paper is as follows: Section \ref{database} describes the detailed information on ECG records and databases which are utilized in this work. Sections \ref{proposed}, \ref{heartrate}, \ref{wavelet} and \ref{ML} describes the proposed multi-class arrhythmia classification methodology followed by hardware implementation of the proposed methodology in Section \ref{pi}. Section \ref{sec:results} discusses all the results obtained from the methodologies presented in the preceding sections. Finally, the conclusion is summarised in Section \ref{conclusion}.

\section{Test Database}\label{database}
\label{sec:Dataset}
The present work uses four open source databases, such as; CPSC database and CPSC-Extra database \cite{b5}, PTB-XL database \cite{b6}, the Georgia 12-lead ECG challenge (G12EC) database and Chapman-Shaoxing and Ningbo database \cite{b7}, \cite{b8},  made available by Physionet for the computing in cardiology challenge 2021 \cite{b4}.

\begin{figure}[!tbh]
	\centerline{\includegraphics[width=\textwidth]{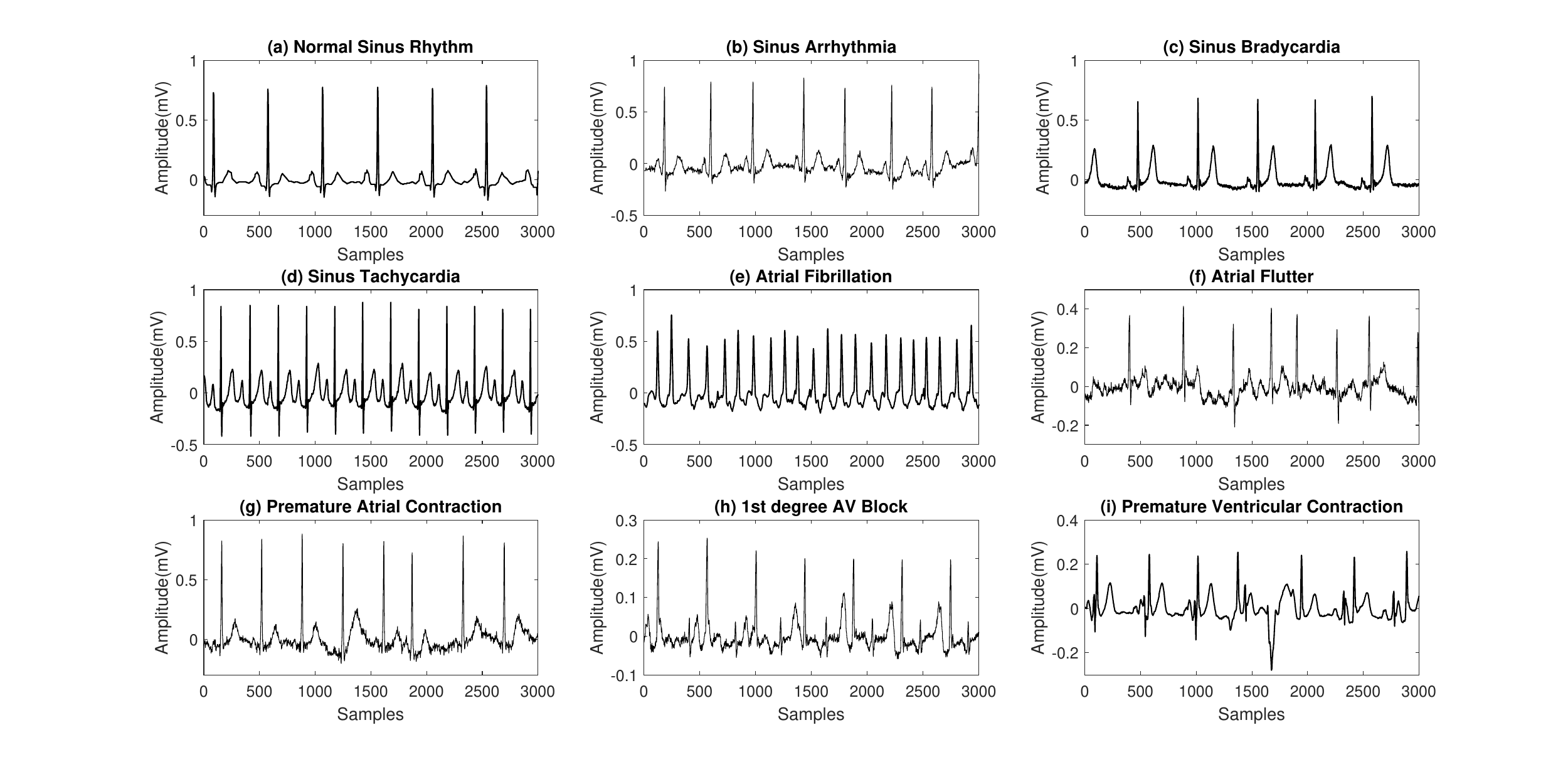}}
	\caption{ECG records having different arrhythmia conditions.}
	\label{ECG images}
\end{figure}

Each database contains a variety of 12-channel ECG records having different arrhythmia conditions. In this work, the lead II ECG signal is utilized for simulation purposes. Each of the signals is sampled at a rate of 500 Hz. The ECG records with PAC and supraventricular premature beats (SVPB) are grouped into a single class of PAC. Similarly, records of PVC and ventricular premature beats (VPB) are merged as the same class PVC. A view of each arrhythmia category of ECG signals is presented in Figure \ref{ECG images}.

The databases are used to accumulate single channel ECG records which form the test dataset. Due to the degradation of the performance of the classifier, the test dataset can be highly imbalanced with 33 records of PVC and 14,993 entries of NSR. Hence, the synthetic minority oversampling technique (SMOTE) and random under-sampling are used to balance the dataset. Initially, SMOTE is applied to increase the data of minority classes \cite{b9}, \cite{sivapalan2022annet}. Further, random under-sampling is  used to reduce the data in the majority class by randomly eliminating the records to obtain a balancing in data \cite{b11}. A total of 31059 ECG records (3451 from each class) are considered in this work. The detailed information on each class of ECG records is described in Table \ref{ECG records in each class}.

\begin{table}
	\centering
	\caption{Number of ECG records of each arrhythmia class}
	\label{ECG records in each class}
	\begin{tabular}{|c|c|c|c|}
		\hline
		\textbf{Arrhythmia class} &\textbf{Original ECG records} & \textbf{After SMOTE} & \textbf{Final ECG records} \\
		\hline
		\textbf{NSR} & 14993 & 14993 & 3451   \\
		\textbf{SA} & 1361 & 3451 & 3451 \\
		\textbf{SB} & 9353 & 9353 & 3451  \\
		\textbf{STACH} & 3451 & 3451 & 3451  \\
		\textbf{AF} & 1500 & 3451 & 3451   \\
		\textbf{AFL} & 1526 & 3451 & 3451  \\
		\textbf{PAC} & 538 & 3451 & 3451  \\
		\textbf{1AVB} & 754 & 3451 & 3451   \\
		\textbf{PVC} & 33 & 3451 & 3451  \\
		\hline
		\multicolumn{4}{p{400pt}}{NSR - Normal sinus rhythm, SA - Sinus Aarrhythmia, SB - Sinus bradycardia, STACH - Sinus tachycardia, AF - Atrial fibrillation, AFL - Atrial flutter, PAC - Premature atrial contraction, 1AVB - 1st Degree AV Block, PVC - Premature ventricular contraction}
	\end{tabular}
\end{table}

\section{Proposed ECG Arrhythmia Classification Methodology}\label{proposed}
\begin{figure}[ht]
    \centerline{\includegraphics[width=0.9\textwidth]{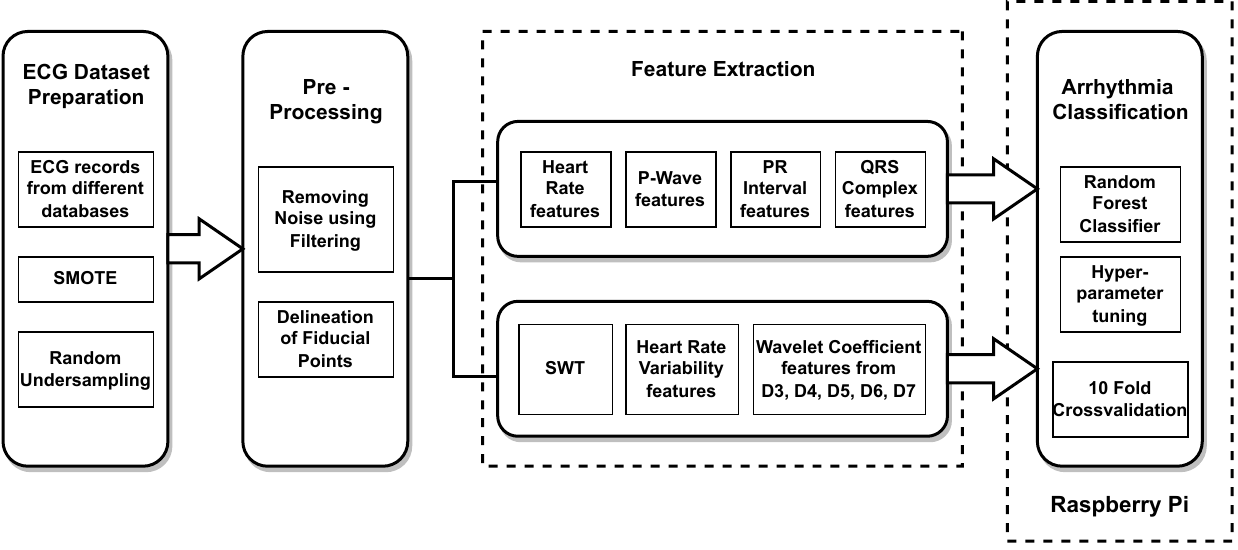}}
     \caption{Flowchart of the proposed arrhythmia detection scheme.}
	\label{workflow}
\end{figure}

The proposed arrhythmia detection and classification methodology is implemented in 3 parts - In the first part, heart rate variability and time domain morphological features from ECG signals are extracted to train a classifier. To improve the results obtained from this classification approach, in the second part, heart rate variability features are paired with wavelet coefficient features obtained from stationary wavelet decomposition of the ECG signal. This gives significantly improved classification performance for each of the 9 classes. In the last part, both the classification methods are implemented in a hardware embedded device to make it viable for real time ECG signal analysis and arrhythmia detection, 

Flowchart of the proposed arrhythmia classification work is presented in Figure \ref{workflow}. The detailed description of each step is discussed in the following subsections.

\subsection{ECG Signal Pre-processing} \label{Preprocessing}
\begin{figure}[h]
	\centerline{\includegraphics[width=\columnwidth]{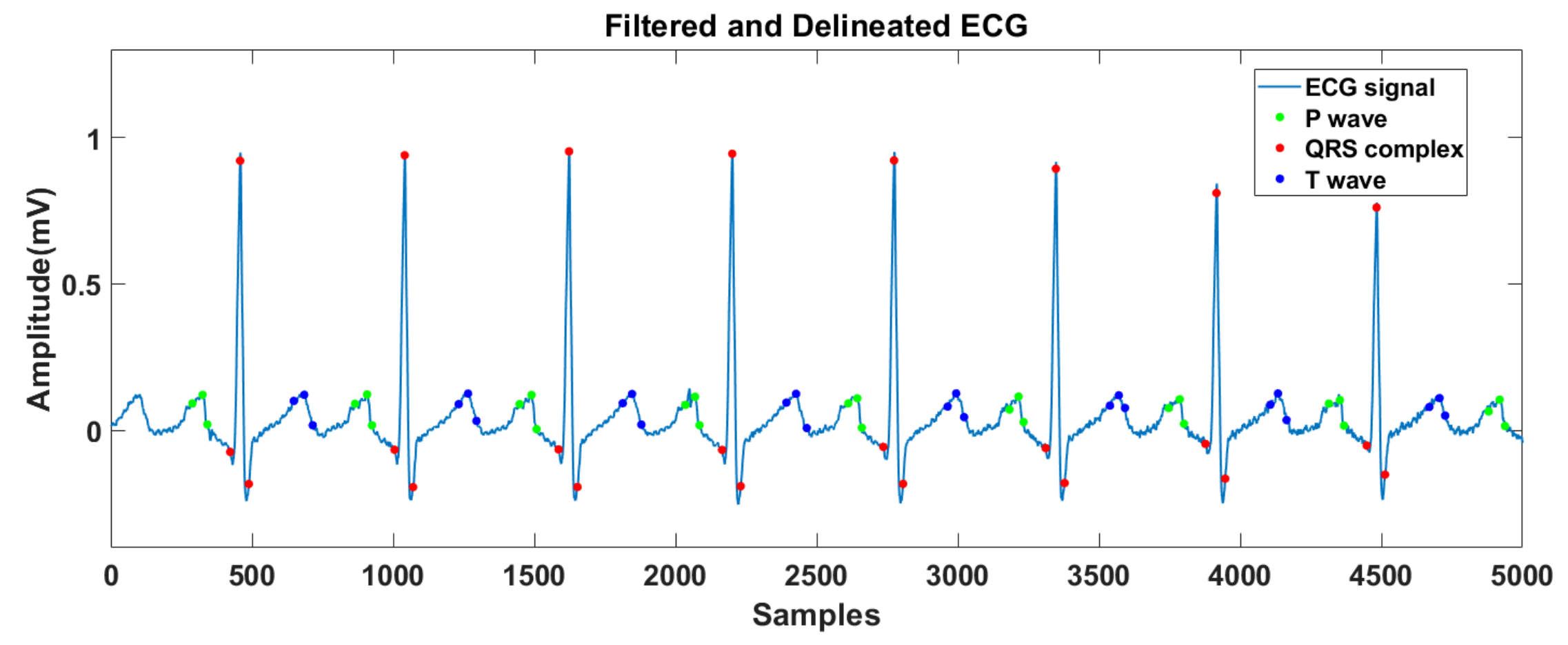}}
	\caption{Local signal component delineation in ECG using ECGDeli toolbox.}
	\label{Filtered ECG}
\end{figure}

The ECG records are often corrupted with baseline wander, muscle artifacts etc. These noises mask the clinical components in the ECG signal which results in poor classification performance \cite{satija2018review}, \cite{chatterjee2020review}. Most of the clinical information of ECG signals is in the frequency range of 1-150 Hz \cite{b3}. For the first part of the work, the ECG signals are filtered through a band-pass filter with a cut-off frequency of 1-150 Hz, which will remove low out-band noise (baseline wander) as well as high frequency noise (muscle artifacts). In the second part, for the wavelet coefficient features, the coefficient sets with frequency range of baseline wander and muscle artifacts are discarded and not used for feature extraction. Powerline interference is removed in both parts of the work using a 50 Hz notch filter. For time domain features, it is crucial to properly delineate the local components (P wave, QRS complex and T wave) in an ECG signal. In this work, ECGDeli, an open-source ECG delineation toolbox is used for accurate delineation of the local components in ECG signals \cite{pilia2021ecgdeli}. The onset, offset and peak of P waves, QRS complex and T waves are reliably detected by this toolbox as shown in Fig \ref{Filtered ECG} 

\section{Heart rate variability and time domain feature extraction}\label{heartrate}
As per the description of arrhythmia in \cite{b3}, it can be concluded that heart rate (HR), regularity of HR, P wave morphology, PR interval (PRI) and QRS complex morphology are the important factors to be considered while looking for the signs of arrhythmia in ECG records. Distinct variations are caused by different types of arrhythmia in HR, P wave morphology, PRI timing and QRS complex morphology. These characteristic variations are analyzed by physicians to detect arrhythmia from ECG recordings and accurate assessment of its type. Hence, in the first part of this work, keeping in consideration the characteristic variations attributed to different types of arrhythmia, detected fiducial points are used to extract the time domain and morphological features which can be grouped under four categories - HR variability features, P wave features, PRI features and QRS complex features. The features are a combination of non-linear, higher-order statistical, entropy and energy-based features so as to best capture the morphological and timing variations in the ECG for different arrhythmia conditions. All categories of features are listed in Table \ref{table2}. The details regarding each category of features are described in the following subsections.

\begin{table}[!h]
		\centering
        \caption{Category wise feature details.}
		\resizebox{17cm}{!}{\begin{tabular}{|l|l|}
		\hline
		&\\
		\Large{\textbf{Feature Category}}&\Large{\textbf{Features}} \\
		\hline                   
		&\\                                                        
		\Large{\textbf{HRV Feature}} & \begin{tabular}[c]{@{}l@{}}\Large{Maximum  HR ($HR_{max}$), Minimum HR ($HR_{min}$), Mean HR ($HR_{mean}$), Std. of HR ($HR_{std}$),} \\ \\ \Large{Max deviation of HR ($HR_{MaxDev}$), Mean of absolute difference of HR ($HR_{MAD}$),} \\\\ \Large{Kurtosis of HR ($HR_{kurt}$), Skewness of HR ($HR_{skew}$), Approximate entropy of HR ($HR_{ApEn}$),} \\\\ \Large{Shannon entropy of HR ($HR_{ShEn}$), Permutation entropy of HR ($HR_{PeEn}$)} \end{tabular} \\&\\ \hline
		
		&\\
		\Large{\textbf{P wave Feature}}      & \begin{tabular}[c]{@{}l@{}} \Large{P wave peak ($P_{peak}$), P wave width ($P_{width}$), Max Deviation of P wave ($P_{MaxDev}$),} \\\\  \Large{P wave energy ($P_{energy}$), Correlation of P waves ($P_{Corr}$), Spectral Entropy of P wave ($P_{SpEn}$),}\\\\ \Large{Kurtosis of P wave ($P_{kurt})$, Skewness of P wave ($P_{skew}$), Atrial HR ($P_{atrialHR}$)} \end{tabular} \\&\\ \hline
		
		&\\
		\Large{\textbf{PRI Feature}}         &    \begin{tabular}[c]{@{}l@{}}\Large{Mean PR Interval ($PRI_{mean}$), Std. of PR Interval ($PRI_{std}$),  Maximum  PR Interval  ($PRI_{max}$),} \\\\ \Large{Minimum  PR Interval ($PRI_{min}$), Max Deviation of PR Interval  ($PRI_{MaxDev}$)} \end{tabular}   \\&\\ \hline
		
		&\\
		\Large{\textbf{QRS Complex Feature}} &   \begin{tabular}[c]{@{}l@{}}\Large{QRS Width  ($QRS_{width}$), Correlation of QRS Complex ($QRS_{Corr}$), QRS complex energy ($QRS_{energy}$)}\\\\ \Large{Spectral Entropy of QRS Complex ($QRS_{SpEn}$), Sample Entropy of QRS ($QRS_{SaEn}$),}\\\\ \Large{Kurtosis of QRS complex ($QRS_{kurt})$, Skewness of QRS complex ($QRS_{skew}$)} \end{tabular}
		\\&\\ \hline 
  \end{tabular}}
	\label{table2}
\end{table}

\subsection{Heart Rate Variability (HRV) Features} \label{HRV1}
The analysis of HR is an imperative step in detecting arrhythmia from ECG signals. Heart rate variation may contain signs of cardiac disease that are already present or warnings of impending cardiac diseases. HR regularity and variability analysis is a crucial step for determining heart rate rhythm as well as arrhythmia \cite{b3}. In this work, a total of 11 features are extracted from heart rate for analyzing the HR regularity and variability. HR is obtained from the interval between two consecutive R peaks. R peaks in this work are readily obtained from ECG deli. If $RR_i$ is the $i^{th}$ RR interval, then for sampling frequency 500 Hz, the HR is calculated as: 
\begin{equation}\label{Heartrate}
	HR_{i} = \frac{60 * 500}{RR_i} \: bpm
\end{equation}

Mean, standard deviation, maximum and minimum HR are some of the linear features calculated. Maximum deviation in HR is a feature calculated as the difference in maximum and minimum HR. From the difference between consecutive HR, the mean absolute difference of HR is also calculated.

\begin{equation}\label{Mean of Abs. Diff}
	HR_{MeanAbsDiff} = \frac{1}{n}\sum_{i=1}^{n}|HR_{i+1}-HR_i|
\end{equation}

Higher order statistics feature such as  kurtosis and skewness of HRV signal are also considered and calculated as follows:

\begin{equation}\label{Kurtosis}
	HR_{Kurt} = \frac{\frac{1}{n}\sum_{i=1}^{n}(HR_i - HR_{mean})^4}{[\frac{1}{n}\sum_{i=1}^{n}(HR_i - HR_{mean})^2]^2}
\end{equation}

\begin{equation}\label{Skewness}
	HR_{Skew} = \frac{\frac{1}{n}\sum_{i=1}^{n}(HR_i - HR_{mean})^3}{[\frac{1}{n}\sum_{i=1}^{n}(HR_i - HR_{mean})^2]^\frac{3}{2}}
\end{equation}
\\

Non-linear entropy based features such as approximate entropy ($HR_{ApEn}$), permutation entropy ($HR_{PeEn}$) and Shannon entropy ($HR_{ShEn}$) are calculated from HRV signal. These features are determined as follows:

For a time series of length $L$, window size of $m$ and threshold of $r$
\begin{equation}\label{ApEn_1}
	C_i^m(r) = \frac{n_{im}(r)}{L+m-1}
\end{equation}
\begin{equation}\label{ApEn_2}
	\emptyset^m(r) = \frac{\sum_{i=1}^{L-m+1}ln[C_i^m(r)]}{L-m+1}
\end{equation}
\\
where, $ C_i^m (r) $ is probability of similarity at threshold $r$, $ n_{im} (r) $ is no. of segments similar to $i^{th}$ segment with threshold $r$ and $ \emptyset^m (r) $ is segment value.\\
Approximate entropy is determined as 
\begin{equation}\label{ApEn_3}
	HR_{ApEn} =  \emptyset^m(r) - \emptyset^{m+1}(r)
\end{equation}
Permutation entropy ($PeEn$) is a measure of complexity taking into account temporal order of the successive points in a time series \cite{b15}. In $PeEn$ calculation total time series is presented into a group of pattern. If $p_{k}$ is probability of occurrence of the $k_{th}$ pattern and $K$ is the length of each pattern then $PeEn$ can be expressed as:
\begin{equation}\label{PeEn}
	HR_{PeEn} = - \sum_{k=1}^{K!}p_klog_2(p_k)
\end{equation}
\begin{equation}\label{Shannon}
	HR_{ShEn} = - \sum_{i=1}^{n}HR_{norm}log_2(HR_{norm})
\end{equation}
where 
\begin{equation}
	HR_{norm} = \frac{HR}{max(HR)}
\end{equation}

\subsection{P wave features}\label{P wave}
P wave in ECG signal is the representation of the atrial activity of the heart and signifies depolarisation of atria. Many of the arrhythmia types (AF, AFL, PAC) affect the morphology of P waves in different ways \cite{b3}. Hence, the morphological information from P wave features can be used effectively to distinguish these arrhythmia classes. Amplitude, width, energy and maximum deviation of P waves are extracted as features which capture essential information like the presence, shape and morphology of P waves. The correlation coefficient of the P wave analyzes the similarity of P waves in consecutive beats. This feature can effectively detect the changes in P wave morphology from beat to beat within an ECG segment in case of an ectopic beat. Other non-linear, statistical features like spectral entropy, skewness, and kurtosis are also calculated for P wave signal components which capture the difference of P wave information for different arrhythmia conditions. Atrial HR is also calculated from the P-P interval which is the time between two consecutive P peaks. As multiple P waves exist throughout the entire ECG segment hence mean and standard deviation are calculated for each of the above-mentioned features. A total group of 18 features are extracted from the P wave signal component.

\subsection{PR interval features}\label{PRI}
PRI is the time duration between the onset of a P wave to the onset of the QRS complex. It signifies the start of atrial depolarisation to the start of ventricular depolarization. It is an important marker for the atrial activity and conduction of electrical impulses through the AV node. The PRI value may be prolonged for the rhythm generated other than sinus or heart block condition. Hence, information on PRI can be used to distinguish between arrhythmia classes. A total 5 PRI features are extracted such as mean PRI ($ PRI_{mean} $), standard deviation of PRI ($ PRI_{std} $), maximum PRI ($ PRI_{max} $), minimum PRI ($ PRI_{min} $) and maximum deviation of PRI ($ PRI_{MaxDev} $).

\subsection{QRS features}\label{QRS}
QRS complex represents the ventricular activity of the heart. The ventricular rhythm arrhythmia like PVC affects the morphology of the QRS complex as compared to normal sinus rhythm. Hence, effective features can be extracted from QRS morphology for better classification. Different morphology, entropy and higher order statics features such as QRS width, correlation coefficient of consecutive QRS, spectral entropy, sample entropy, kurtosis and skewness are extracted from QRS complexes. Similar to P waves multiple QRS complexes exist throughout the entire ECG signal hence mean and standard deviation are calculated for each of the above-mentioned features. A total group of 14 features are extracted from QRS complexes.

\section{Wavelet coefficient feature extraction}\label{wavelet}

To improve the methodology proposed in the first part, in the second part, an arrhythmia detection scheme is devised where in addition to heart rate features, wavelet domain features are obtained from coefficients of wavelet decomposition of the ECG signals using SWT. 

\subsection{SWT based signal decomposition}\label{SWT}
Wavelet transform is a signal processing tool that allows the decomposition of the signal into different time and frequency scales where each scale allows analysis of various signal properties and characteristics. This tool for analyzing non-stationary signals is useful and simple for identifying subtle variations in the signal morphology over the scales of interest \cite{b26}. A series of high and low-pass filters are used to analyze high and low-frequency components of the signal. At each level, convolution of the input signal with high pass filters gives the detail coefficients $D_n$ and convolution with low pass filters gives approximation coefficients  $A_n$. 

SWT is a time-invariant discrete wavelet transform method. At each decomposition level, SWT coefficients have the same number of samples as the original signal, thus preserving the temporal information of the signal and overcoming the problem of repeatability and robustness which exists with discrete wavelet transform\cite{b26}. SWT of a signal $x[n]$ gives the coefficients $c_{i,j}$ obtained using equation \cite{merah2015r} - 

\begin{equation}\label{Transform Coefficient}
	c_{i,j} = \sum_{n \in z}x[n]\psi^*_{i,j}[n]
\end{equation}
where $\psi_{i,j}$ is a translated and scaled version of the mother wavelet $\psi_{0,0}$
\begin{equation}\label{Mother Wavelet}
	\psi_{i,j}[k] = 2^{-(i/2)}\psi_{0,0}(2^{-i}(k-j))
\end{equation}

To implement $L$-level SWT on a signal, the length of the signal should be a multiple of $2^L$. Signals with lesser samples are zero-padded to make the signal length a multiple of $2^L$. At each successive decomposition level, the impulse response of the high and low pass filters are upsampled by a factor of 2 giving a coefficient series with the same temporal resolution as the original signal.

\begin{figure}[!t]	\centerline{\includegraphics[width=0.90\columnwidth]{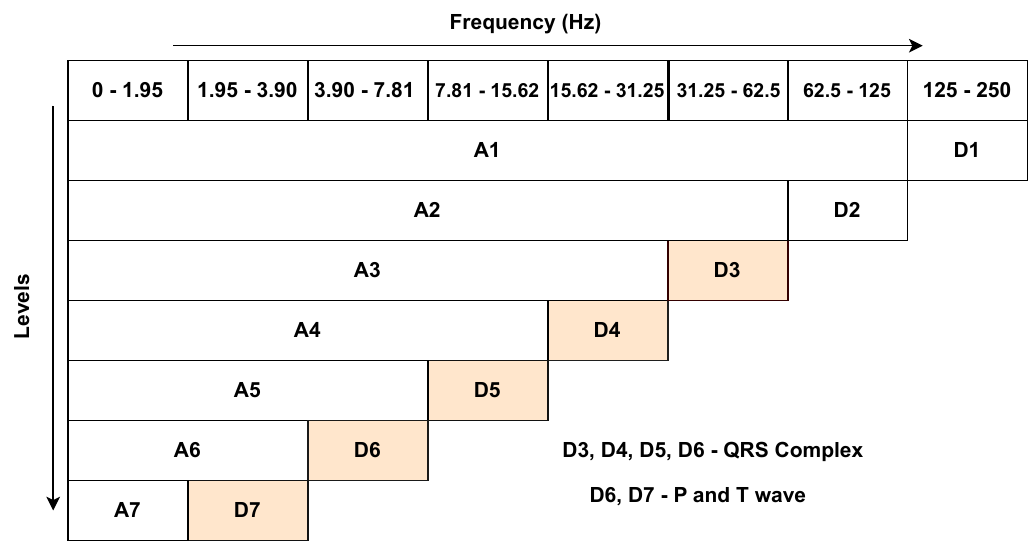}}
	\caption{Frequency range information of SWT decomposition sub-band levels.}
	\label{Freq Decomposition}
\end{figure}

In this work, the ECG records are sampled at 500 Hz with a maximum frequency component of 250 Hz. Considering the frequency ranges of the QRS complex, P wave and T wave, a 7-level SWT decomposition is chosen to be applied to the ECG records using Symlet-7 wavelet. Symlet-7 wavelet is chosen as the mother wavelet because of its close similarity to ECG signal morphology and is extensively used in different ECG signal processing-based works \cite{ansari2017review}. The frequency range of each decomposition level is shown in Figure \ref{Freq Decomposition}. Detail coefficients D3, D4, D5 and D6 correspond to the frequency range of the QRS complex while D6 and D7 correspond to the P and T waves of the ECG signal. Thus, D3, D4, D5, D6 and D7 are considered for wavelet coefficient feature calculation in the subsequent steps of the algorithm. 

\begin{figure}[!t]	\centerline{\includegraphics[width=\columnwidth]{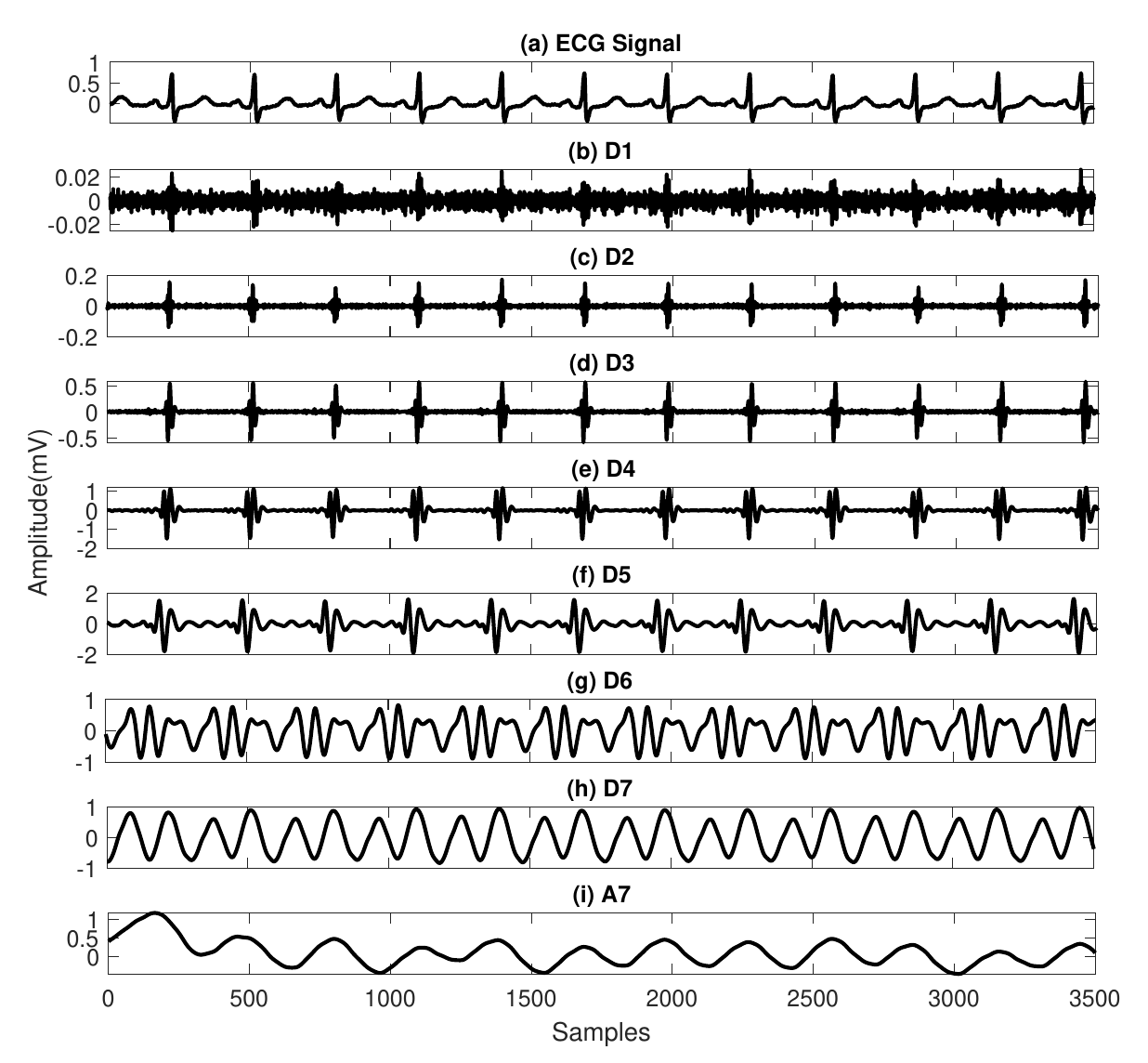}}
	\caption{Original ECG with decomposed signals (a) Original ECG ; (b)-(h) Detail coefficients D1 - D7 ; (i) Approximate coefficient A7.}
	\label{SWT Decomposition plot}
\end{figure}

\subsection{HRV features}\label{HRV2}
In addition to the 11 HR features from the Subsection \ref{HRV1}, 5 more features were added in this scheme to increase the arrhythmia classification accuracy as HR features contribute the most in differentiating multiple classes of arrhythmia. These new features are - standard deviation of absolute difference, coefficient of variance (CoV) of HR, Higuchi fractal dimension, Hjorth mobility and Hjorth complexity. CoV of HR is the ratio of the standard deviation of HR to the mean of HR.
Higuchi fractal dimension is a computational method to determine changes in a signal from the measure of its complexity \cite{b29}. For a finite set of time series observations taken at regular intervals $X(1), X(2), X(3)....X(N)$, a new time sub series can be constructed by taking
\begin{equation}
	X_k^m : X(m), X(m+k), X(m+2k),..., X(m+[\frac{N-m}{k}].k)
\end{equation}
where $m = 1,2,3,..,k$ is the start of every sub series and $k$ is the interval. This gives a total of $k$ new sub series. The length of each of the sub series is given by
\begin{equation}
	L_m(k) = \frac{{\lfloor{\sum_{i=1}^{[\frac{N-m}{k}]}x_{m+id}-x_{m+(i-1)d}}}\rfloor\frac{N-1}{\frac{N-m}{k}.k}}{k}
\end{equation}
On plotting the average value of the length over $k$ sub series against $k$ in a double logarithmic scale, the data falls on a straight line with slope-D. The value $D$ is Higuchi fractal dimension for the given time series \cite{b30}.

Hjorth mobility ($HR_{HM}$) is a Hjorth descriptor that describes the mean frequency of a signal. Hjorth complexity ($HR_{HC}$) denotes an estimation of the signal bandwidth from the ratio of the peak value to the harmonic content of the signal \cite{b31}. These two features are calculated as:

\begin{equation}
	HR_{HM} = \frac{\sigma'_x}{\sigma_x}
\end{equation}

\begin{equation}
	HR_{HC} = \frac{\frac{\sigma''_x}{\sigma'_x}}{\frac{\sigma'_x}{\sigma_x}}
\end{equation}

where, $\sigma_x$ is the variance of the time series, $\sigma'_x$ is the 1st derivative of the variance and $\sigma''_x$ is the $2^{nd}$ derivative of the variance.

\subsection{Wavelet coefficient features}\label{Wavelet Coeff Features}
Detail coefficients D3, D4, D5, D6 and D7 correspond to the frequency range of clinically significant information in ECG signal. 10 features are extracted from each of these four sets of coefficients to give a total of 50 features. The feature set includes statistical, entropy and energy based features. Statistical features such as mean, standard deviation, skewness and kurtosis are obtained from each set of detail coefficients. Entropy features of approximate entropy, Shannon entropy, permutation entropy and log energy entropy ($LEEn$), when applied to wavelet coefficients, capture the complexity, regularity and uncertainty in the wavelet decomposed subbands of the ECG signal. Two energy based features namely: relative wavelet energy ($RWE$) and mean wavelet energy ($MWE$) are also considered. The calculation of $LEEn$, $RWE$ and $MWE$ are as follows \cite{b32}:  

\begin{equation}
	LE_{En} = \sum_{i=1}^Nlog(x_i^2)
\end{equation}
where, $x_j$ is the $i^{th}$ sample and $N$ is the length of the sub-band signal.
\begin{equation}
	RWE = \frac{\sum_{i=1}^{N}C_j(i)^2}{\sum_{j=1}^{L}\sum_{i=1}^{N}C_j(i)^2}
\end{equation}

\begin{equation}
	MWE = \frac{\sum_{i=1}^{N}C_j(i)^2}{N}
\end{equation}
where, $N$ is the total number of coefficients in $j^{th}$ level and $L$ is the total number of decomposition levels.

\begin{table} [!htb]
	\centering
	\caption{Set of wavelet coefficient features}
	\begin{tabular}[h]{|p{5cm}|}
		\hline
		\textbf{Wavelet features}\\ 
        \hline
        Mean ($CD_{mean}$)\\
        Standard deviation ($CD_{std}$)\\
        Skewness ($CD_{skew}$) \\
        Kurtosis ($CD_{kurt}$) \\
        Approximate entropy ($CD_{ApEn}$) \\
        Shannon entropy ($CD_{ShEn}$) \\
        Permutation entropy  ($CD_{PeEn}$)\\
        Log energy entropy ($CD_{LE_{En}}$) \\
        Relative wavelet energy  ($CD_{MWE}$) \\
        Mean wavelet energy  ($CD_{RWE}$)\\
		\hline
	\end{tabular}
	
	\label{tab:Wavelet features }
\end{table}

\section{Machine Learning based Arrhythmia Classifier}\label{ML}
In this section, a machine learning based classifier is utilized for the automated detection of arrhythmia. The supervised machine learning based approaches require feature matrix along with labels. In part one of the work, a total of 48 features (11 HRV, 18 P wave, 5 PRI and 14 QRS features) made up the feature set and in next part, there were 66 total features (16 HRV and 50 wavelet features) in the feature set. Feature sets obtained in the previous sections are supplied to the classifier. There are several machine learning-based techniques and each of them is extensively used in many literature. Random forest (RF) is one of the popular ensemble classifiers. It is extensively employed in many classification problem including medical signal and image processing \cite{kung2020efficient}, \cite{panayides2020ai}. RF is an ensemble classier which is made from bootstrap aggregation of multiple decision trees. Each decision tree independently generates an output as per the input feature matrix supplied to it. Finally, the net resultant output can be found by applying a voting strategy on the outputs from multiple trees. The aggregation of voting makes RF more effective classifier and less susceptible to outliers and noises \cite{shaikhina2019decision}. A comparative analysis on different machine learning classifiers is studied in the following subsection and it can be checked that the RF performs better compared to others. Hence, RF is used here as the arrhythmia classifier. Hyper-parameter tuning is done separately for both features sets to obtain a set of optimal parameters for the RF classifier giving the best performance.  

\section{Hardware implementation of the proposed classification schemes}\label{pi}
To validate the proposed arrhythmia detection and classification schemes, this section presents the hardware implementation using a Raspberry Pi 4 model B. A programmable system on a chip like Raspberry Pi can handle the complexity of computation while keeping the power consumption low. Raspberry Pi is a single board computer working on a Linux based operating system and takes real time input data which can then be used for a multitude of applications. Apart from this low cost Pi, a computer for feature extraction from dataset and a display monitor to view the output from the Raspberry Pi is used as hardware components. The ECG dataset consisting of 31,059 10-sec ECG records is given as input to the feature extraction algorithm. Feature extraction in both the proposed schemes is done using Matlab 2022b. The obtained feature set is given as input to the Raspberry Pi where the classifier is deployed. The classifier model is written in Python programming language using scikit-learn library. The classifier is trained on the feature set and the trained model then can be used to detect and classify arrhythmic rhythms in real time ECG signals.

\begin{figure}[h]
\centerline{\includegraphics[width=0.90\columnwidth]{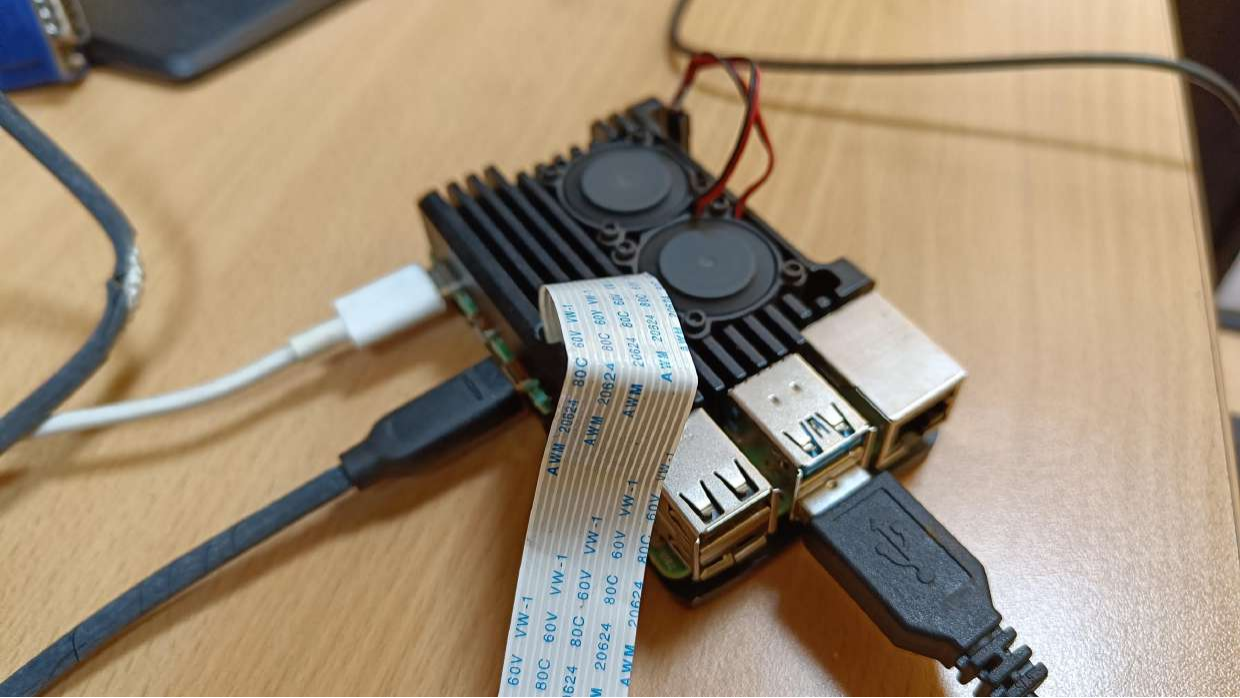}}
	\caption{Hardware implementation with Raspberry Pi 4B}
	\label{R Pi connection}
\end{figure}

\begin{figure}[h]
\centerline{\includegraphics[width=0.90\columnwidth]{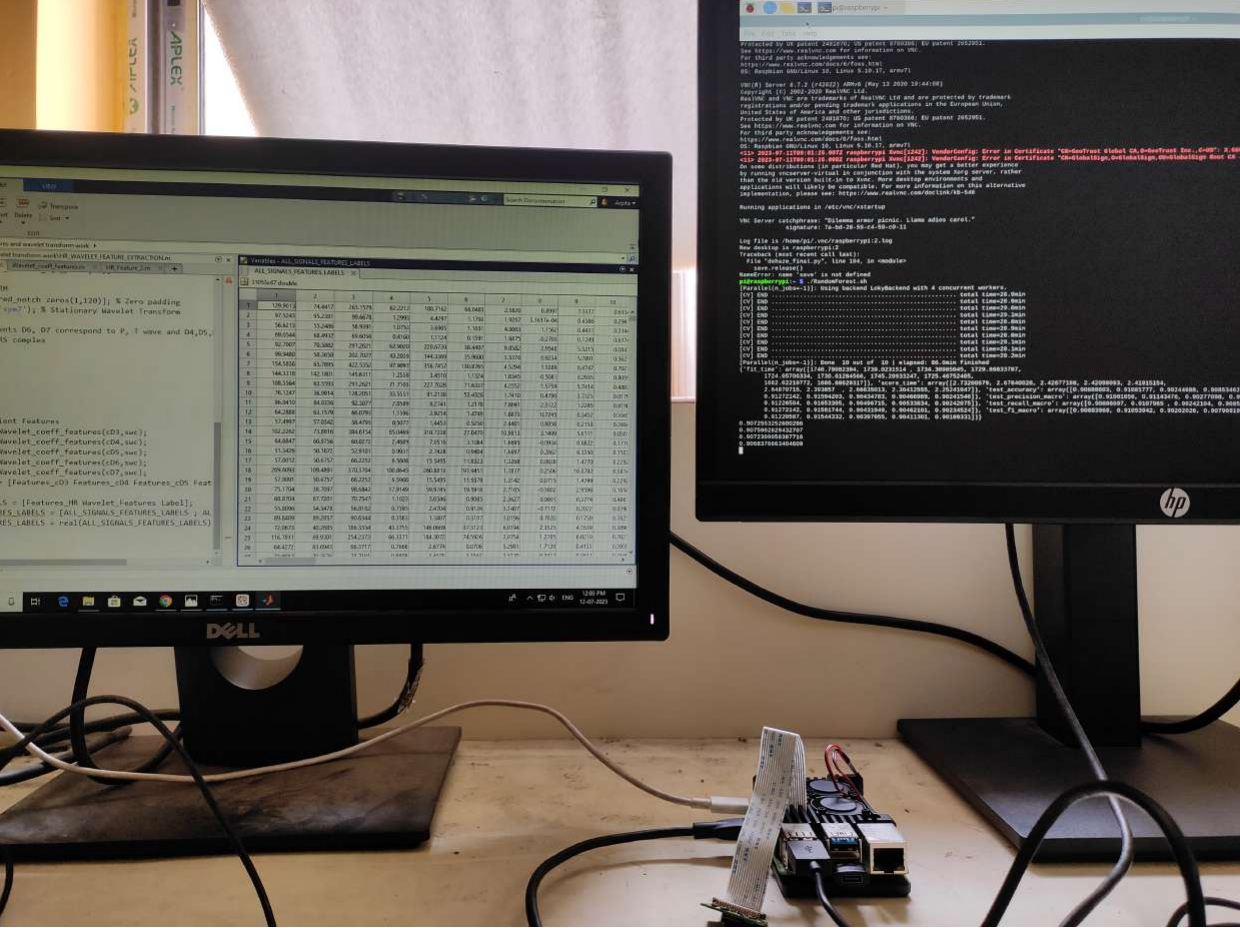}}
	\caption{Feature extraction is fed to Raspberry Pi 4B (left) for feature extraction and a display monitor (right) to show output obtained from classifier loaded on Raspberry Pi 4B}
	\label{R Pi output}
\end{figure}

\section{Results and Discussion}
\label{sec:results}
Performance of both the proposed cardiac arrhythmia detection scheme is presented in this section. Training and testing of the classifier is done through 10-fold cross-validation. In 10-fold cross-validation, the entire dataset is divided into 10 subsets. At each time, one subset is used for testing purposes and the remaining nine subsets are used for training the algorithm. This process is repeated for all 10 sub-sets and finally, the average values of the performance metrics are reported. The performance of the proposed arrhythmia detection method is evaluated using standard metrics: accuracy ($Acc$), sensitivity ($Se$), precision ($+P$) and $F1$ score. These are calculated from parameters: true positives ($TP$), true negative ($TN$), false positive ($FP$) and false negative ($FN$).

\begin{equation}\label{Accuracy}
	Accuracy (Acc) = \frac{TP + TN}{TP + TN + FP + FN}
\end{equation}

\begin{equation}\label{Sensitivity}
	Sensitivity (Se) = \frac{TP}{TP + FN}
\end{equation}

\begin{equation}\label{Precision}
	Precision (+P) = \frac{TP}{TP + FP}
\end{equation}

\begin{equation}\label{F1}
	F1 Score = \frac{2*TP}{2*TP + FP + FN} 
\end{equation}
\\

The detailed 10-fold cross-validation performance result of classification with HRV, P wave, PRI and QRS complex morphological features are presented in Table \ref{CV Performance result 1} . Average performance obtained is $Acc$ of 85.11\%, $Se$ of 85.11\%, $P$ of 85.07\% and $ F1 $ score of 85\%.  

\begin{table} [h]
	\centering
	\caption{10- fold Cross-validation performance results with heart rate and time domain features}
	\begin{tabular}[h]{|c|c|c|c|c|}
		\hline
		\textbf{Fold } & \textbf{Acc(\%)} & \textbf{Se(\%)} & \textbf{+P(\%)} & \textbf{F1(\%)}\\
		\hline 
		1 & 85.64 & 85.64 & 85.65 & 85.57  \\
		2 & 85.12  & 85.12 & 85.02 & 85.00 \\
		3 & 84.86 & 84.87 & 84.81 & 84.77\\
		4 & 84.83 & 84.84 & 84.79 & 84.68\\
		5 & 85.57 & 85.57 & 85.55 & 85.46\\
		6 & 85.47 & 85.47 & 85.31 & 85.26\\
		7 & 85.44 & 85.44 & 85.46 & 85.34\\
		8 & 85.02 & 85.02 & 84.98 & 84.93\\
		9 & 84.93 & 84.92 & 84.95 & 84.86\\
		10 & 84.21 & 84.21 & 84.15 & 84.09\\
		\textbf{Average} & \textbf{85.11} & \textbf{85.11} & \textbf{85.07} & \textbf{85.00}\\
		\hline
	\end{tabular}
	\label{CV Performance result 1}
\end{table}

Class-wise classification performance of the described method is presented in Table \ref{Classwise performance 1}. Maximum classification $ F1 $ score of 96.36\% is obtained for STACH. Lowest $ F1 $ score of 75.31\% and 71.60\% is obtained for the classes of AF and AFL.

\begin{table} [h]
	\centering
	\caption{Class wise performance of arrhythmia classification with heart rate and time domain features}
	%\begin{tabular}[h]{|m{1cm}|m{1cm}|m{1cm}|m{1cm}|}
        \begin{tabular}[h]{|c|c|c|c|}
		\hline
		\textbf{Class} & \textbf{Se(\%)} & \textbf{+P(\%)} & \textbf{F1(\%)}\\
		\hline 
		\textbf{NSR} & 79.98 & 92.00 & 85.57  \\
 	\textbf{SA}  & 86.58 & 86.56 & 86.57 \\
 	\textbf{SB}   & 97.19 & 89.04 & 92.93\\
 	\textbf{STACH}  & 97.74 & 95.01 & 96.36\\
		\textbf{AF} & 75.02 & 75.59 & 75.31\\
 	\textbf{AFL}   & 70.65 & 72.58 & 71.60\\
 	\textbf{PAC}   & 79.83 & 79.76 & 79.80\\
 	\textbf{1AVB}   & 83.14 & 83.74 & 83.44\\
 	\textbf{PVC}   & 96.35 & 91.70 &93.97 \\
		\hline
	\end{tabular}
	\label{Classwise performance 1}
\end{table}

To have a more in depth analysis of the feature set with HRV features and three categories of time domain and morphological features, performance of different subsets of the feature set is compared in Table \ref{Feature subset performance comparision 1}. It can be observed that the HRV feature has the superior effectiveness of detecting arrhythmia with a maximum classification $ Acc $ of 82.81\%. It solidifies the importance of HR features for arrhythmia classification. PRI features alone are the least accurate in multi-class arrhythmia classification as the PRI segment mostly signifies the conduction time of impulse through the AV node. Similarly, P wave features and QRS complex features also provide less classification accuracy compared to the HRV feature set when used individually. Adding P-wave, PRI and QRS complex features along with  HRV features makes the classification of nine classes more robust and accurate. It is also established that each of the four category features adds more robustness to the algorithm and makes it efficient in multi-class arrhythmia classification. This gives important insights into the clinical relevance of different ECG signal components.

\begin{table} [ht]
	\centering
        \caption{Performance comparison with different subsets of HRV and time domain features.}
	\begin{tabular}[h]{|c|c|c|c|c|}
		\hline
		\textbf{Feature Category} & \textbf{Acc(\%)} & \textbf{Se(\%)} & \textbf{P(\%)} & \textbf{F1(\%)}\\
        \hline
		\textbf{HRV}  & \textbf{82.81} & \textbf{82.81} & \textbf{82.64} & \textbf{82.6}1  \\
		\textbf{P Wave} & 73.10  & 73.10 & 72.87 & 72.69 \\
		\textbf{PRI} & 43.65  & 43.65 & 42.83 & 43.06 \\
		\textbf{QRS Complex} & 59.19 & 59.19 & 59.58 & 58.81 \\
		\textbf{HRV + P Wave} & 80.48 & 80.48 & 80.18 & 80.15 \\
		\textbf{HRV + P Wave + PRI} & 82.46 & 82.46 & 82.43 & 82.33 \\
		\textbf{HRV + P Wave + PRI + QRS} & \textbf{85.11} & \textbf{85.11} & \textbf{85.07} & \textbf{85.00}   \\
		\hline
	\end{tabular}  
	\label{Feature subset performance comparision 1}
\end{table}

Next, results of classification with HRV and wavelet features are presented in Table \ref{tab:CV Performance result 2 }. 10-fold cross-validation with the proposed scheme gives an average $Acc$ of 90.91\% , $Se$ of 90.91\%, $+P$ of 90.96\% and $F1$ score of 90.87\%. Both the methods show similar results for each validation subset which suggests the robust classification performance of the proposed schemes. 

\begin{table} [!htb]
	\centering
	\caption{Detailed cross-validation results with heart rate and wavelet features}
	\begin{tabular}[h]{|c|c|c|c|c|}
		\hline
		\textbf{Fold} & \textbf{Acc(\%)} & \textbf{Se(\%)} & +\textbf{P(\%)} & \textbf{F1(\%)} \\
		\hline 
		1 & 90.98 & 90.98 & 91.09 & 90.98\\
		2 & 91.21 & 91.21 & 91.25 & 91.18\\
		3 & 90.18 & 90.18 & 90.20 & 90.13\\
		4 & 91.07 & 91.07 & 91.05 & 91.02\\
		5 & 90.62 & 90.62 & 90.66 & 90.59\\
		6 & 91.36 & 91.37 & 91.34 & 91.32\\
		7 & 91.69 & 91.69 & 91.80 & 91.66\\
		8 & 90.62 & 90.62 & 90.68 & 90.59\\
		9 & 90.59 & 90.59 & 90.67 &  90.53\\
		10 & 90.78 & 90.78 & 90.80 & 90.73\\
		\textbf{Average} & \textbf{90.91} & \textbf{90.91} & \textbf{90.96} & \textbf{90.87} \\
		\hline
	\end{tabular}
	
	\label{tab:CV Performance result 2 }
\end{table}

The performance of the proposed scheme for individual arrhythmia class is described in Table \ref{tab:Class wise performance result }. The method shows a maximum classification $F1$ score of 98.00\% for PVC. The $F1$  score of AFL arrhythmia class is marginally small at 80.81\%  compared to other classes. The ECG morphology characteristics of AF and AFL arrhythmia classes are quite similar, hence for these two classes, both the methods give larger false detection compared to other classes. This results in a degradation of performance for AFL arrhythmia classification.

\begin{table} [!htb]
	\centering
	\caption{Class-wise performance result with heart rate and wavelet features}
	\begin{tabular}[h]{|c|c|c|c|}
		\hline
		\textbf{Class} & \textbf{Se(\%) }& \textbf{+P(\%)} & \textbf{F1(\%)} \\
		\hline 
		\textbf{NSR} & 87.54 & 92.08 & 89.75  \\
		\textbf{SA}  & 91.97 & 88.09 & 89.99 \\
		\textbf{SB}   & 97.74 & 93.07 & 95.35\\
		\textbf{STACH}  & 98.12 & 95.49 & 96.78\\
		\textbf{AF} & 79.27 & 87.11 & 83.01\\
		\textbf{AFL}   & 82.61 & 79.08 & 80.81\\
		\textbf{PAC}   & 88.03 & 90.23 & 89.12\\
		\textbf{1AVB}   & 85.71 & 88.80 & 87.23 \\
		\textbf{PVC}   & 99.57 & 96.49 & 98.00 \\
		\hline
	\end{tabular}
	
	\label{tab:Class wise performance result }
\end{table}

Proper selection of the mother wavelet in SWT-based decomposition is a crucial task. Generally, a wavelet is chosen whose shape is nearly similar to the morphology of the ECG cycle. Different mother wavelets such as Daubechies, Coiflet, Symlet, Bi-orthogonal etc are extensively used in several ECG signal processing-related works. In this work, a comparative performance study has been carried out using different mother wavelets. As presented in Table \ref{tab:Wavelet Performance comparison}, the Symlet wavelet of order 7 (Symlet 7) shows a better performance of accuracy of 90.91\%.  

\begin{table} [!htb]
	\centering
	\caption{Performance comparison of the proposed scheme for different mother wavelets}
	\begin{tabular}[h]{|c|c|c|c|c|}
		\hline
		\textbf{Wavelet} & \textbf{Acc(\%)} & \textbf{Se(\%)} & \textbf{+P(\%)} & \textbf{F1(\%)} \\
		\hline
		\textbf{Haar} & 89.97 & 89.97 & 89.98 & 89.90 \\
		\textbf{Daubechies 6} & 90.41 & 90.41 & 90.40 & 90.35 \\
		\textbf{Coiflet 2} & 90.86 & 90.86 & 90.89 & 90.82 \\
		\textbf{Biorthogonal 4.4} & 90.90 & 90.90 & 90.93 & 90.86 \\
		\textbf{Symlet 7} & \textbf{90.91} & \textbf{90.91} & \textbf{90.96} & \textbf{90.87} \\
		
		\hline
	\end{tabular}
	
	\label{tab:Wavelet Performance comparison}
\end{table}

The arrhythmia classification performance of HRV features along with different combinations of wavelet coefficient features is presented in Table \ref{tab:Feature Category Performance comparison}. It can be observed that the HRV features alone have an effective classification accuracy of 82.81\%. Wavelet features of D3, D4, D5, D6 and D7 coefficient set give an accuracy of 83.04\% without including HRV features. Performance gradually increases on combining the HRV features and wavelet coefficient feature sets. This justifies the effectiveness of HRV features along with wavelet coefficient features for the detection of multi-class cardiac arrhythmia. 

\begin{table} [!htb]
	\centering
	\caption{Performance comparison of proposed scheme for different feature combinations}
	\begin{tabular}[h]{|c|c|c|c|c|}
		\hline
		\textbf{Feature Category} &\textbf{Acc(\%)} &\textbf{Se(\%) }& \textbf{+P(\%)} & \textbf{F1(\%)} \\
		\hline
		\textbf{HRV} & 82.81 & 82.81 & 82.64 & 82.61 \\
        \textbf{D3 + D4 + D5 + D6 + D7} & 83.04 & 83.03 & 82.88 & 82.73 \\
        \textbf{HRV + D3} & 89.45 & 89.45 & 89.43 & 89.40 \\
		\textbf{HRV + D3 + D4} & 89.57 & 89.57 & 89.54 & 89.52 \\
		\textbf{HRV + D3 + D4 + D5} & 89.62 & 89.62 & 89.63 & 89.58 \\
		\textbf{HRV + D3 + D4 + D5 + D6} & 90.45 & 90.45 & 90.47 & 90.41 \\
		\textbf{HRV+ D3 + D4 + D5 + D6 + D7} & \textbf{90.91} & \textbf{90.91} & \textbf{90.96} & \textbf{90.87} \\
		\hline
	\end{tabular}  
	\label{tab:Feature Category Performance comparison}
\end{table}

The machine learning-based classifier for arrhythmia detection is an imperative component in this work. Output of the classification is highly dependent on the proper selection of a machine learning classifier. In this section, a comparative analysis is studied on the performance of both schemes for different machine learning classifiers. The techniques are tested on three other extensively used classifiers namely: K- nearest neighbours (KNN), decision tree (DT), support vector machine (SVM) and RF. As presented in Table \ref{tab:Classifier Performance comparison}, maximum accuracy is obtained with RF classifier. Aggregation of the voting concept in RF improves the classification performance compared to other classifiers.

\begin{table*}[!htb]
  \centering
  \caption{Performance comparison with different machine learning classifiers}
  \begin{tabular}{|l|l|l|l|l|l|l|l|l|}
    \hline
    \multirow{2}{*}{\textbf{Classifier}} &
      \multicolumn{4}{|c|}{\textbf{HRV and time domain features}} &
      \multicolumn{4}{|c|}{\textbf{HRV and wavelet features}} \\ \cline{2-9}
    & Acc(\%) & Se(\%) & +P(\%) & F1(\%) & Acc(\%) & Se(\%) & +P(\%) & F1(\%) \\
    \hline
   \textbf{KNN} & 72.97 & 72.97 & 73.01 & 72.77 & 79.29 & 79.28 & 79.16 & 79.08 \\
    \textbf{Decision Tree} & 74.22  & 74.22 & 74.29 & 74.23 & 81.76 & 81.76 & 81.80 & 81.75\\
    \textbf{SVM} &  81.30 & 81.30 & 81.21 & 81.13 & 85.76 & 85.76 & 85.81 & 85.64\\
    \textbf{Random Forest} & \textbf{85.11} & \textbf{85.11} & \textbf{85.07} & \textbf{85.00} & \textbf{90.91} & \textbf{90.91} & \textbf{90.96} & \textbf{90.87} \\
    \hline
  \end{tabular}
  \label{tab:Classifier Performance comparison}
\end{table*}

\section{Conclusion}\label{conclusion}

In this work, multi-class cardiac arrhythmia detection schemes are proposed. In the first part, HRV features are incorporated together with time domain statistical, entropy and higher order statistical features of P wave, PR interval and QRS complex for the effective classification of cardiac arrhythmia using single-channel ECG records. A set of total 48 features are applied to a machine learning-based random forest classifier. The detailed 10-fold cross-validation results show that the proposed multi-class cardiac arrhythmia detection algorithm effectively classifies nine rhythms with an average Acc of 85.11\%, Se of 85.11\%, P of 85.07\% and F1 score of 85.00\%. In the second part, wavelet coefficient features are used along with HRV features to successfully classify different arrhythmia types. Initially, stationary wavelet transform is applied to decompose the ECG signal into different sub-band levels. Considering the frequency localization property of wavelet transform, matching with the frequency of ECG local components, the detail coefficients D3, D4, D5, D6, and D7 are further processed for feature extraction. A set of 66 features based on timing information, entropy, higher-order statistics, and energy are extracted and applied to RF classifier. Detailed cross-validation results show that the proposed multi-class cardiac arrhythmia detection algorithm can effectively classify nine rhythm type with average Acc of 90.91\%, Se of 90.91\%, +P of 90.96\% and F1 score of 90.87\%. For both parts of the work, ECG records of broadly distributed four standard databases of the Physionet Challenge 2021 are combined to prepare a test database having nine classes of arrhythmia. Lastly, both the classification schemes are implemented on Raspberry Pi. It's low power consumption, light weight and compact design makes it suitable for application in real time monitoring and processing of ECG signals. A close observation of the simulation results affirm that the proposed schemes can effectively be utilized in an advanced automated cardiac disease monitoring system.

\section{Acknowledgment}

The authors would like to thank Dr. Ankita Pramanik, Assistant Professor, Electronics and Telecommunication Engineering Department, IIEST, Shibpur, for her guidance and support.

\bibliographystyle{unsrtnat}
\bibliography{MyRef}  %%% Uncomment this line and comment out the ``thebibliography'' section below to use the external .bib file (using bibtex) .

%%% Uncomment this section and comment out the \bibliography{references} line above to use inline references.
% \begin{thebibliography}{1}

% 	\bibitem{kour2014real}
% 	George Kour and Raid Saabne.
% 	\newblock Real-time segmentation of on-line handwritten arabic script.
% 	\newblock In {\em Frontiers in Handwriting Recognition (ICFHR), 2014 14th
% 			International Conference on}, pages 417--422. IEEE, 2014.

% 	\bibitem{kour2014fast}
% 	George Kour and Raid Saabne.
% 	\newblock Fast classification of handwritten on-line arabic characters.
% 	\newblock In {\em Soft Computing and Pattern Recognition (SoCPaR), 2014 6th
% 			International Conference of}, pages 312--318. IEEE, 2014.

% 	\bibitem{hadash2018estimate}
% 	Guy Hadash, Einat Kermany, Boaz Carmeli, Ofer Lavi, George Kour, and Alon
% 	Jacovi.
% 	\newblock Estimate and replace: A novel approach to integrating deep neural
% 	networks with existing applications.
% 	\newblock {\em arXiv preprint arXiv:1804.09028}, 2018.

% \end{thebibliography}

\end{document}